# INFECTION IN A CONFINED SPACE USING AN AGENT- BASED MODEL

**Sarumi, J. A (PhD)[1], Onwubiko, E. C.[2], Ogunjimi O. L. A.(PhD, FA**
jerrytechnologies@yahoo.co.uk[1], econwubiko@gmail.com[2], olaogunjimi@gmail.com

**Lagos State Polythetic Ikorodu, Lagos State[1]; Advance Aerospace Engineering Laboratory, Oka Akoko, Ondo State[2]; Augustine University Ilara-Epe, Lagos State[3].**

## Abstract

As the world observes a new pandemic with COVID-19, it is clear that pathogens can spread rapidly and without recognition of borders. Outbreaks will continue to occur, and so the diseases' transmission method must be thoroughly understood in order to minimize their impact. Some infections, such as influenza, tuberculosis and measles are known to be spread through droplets in the air. In a confined space the concentration can grow as more droplets are released. This study examined a simulated confined space modelled as a hospital waiting area, where people who could have underlying conditions congregate and mix with potentially infectious individuals. It further investigated the impact of the volume of the waiting area, the number of people in the room, the placement of them as well as their weight. The simulation is an agent-based model (ABM), a computational model with the purpose of analyzing a system through the actions and cumulative consequences of autonomous agents. The presented ABM features embodied agents with differing body weights that can move, breathe and cough in a ventilated room. An investigation into current epidemiological models lead to the hypothesis that one may be implemented as a corresponding ABM, where it could possi- bly also be improved upon. In this paper, it is shown that all parameters of the Gammaitoni and Nucci model can be taken into account in an ABM via the MASON library. In addition, proof is produced to suggest that some flaws of the epidemiological model can be mended in the ABM. It is demonstrated that the constructed model can account for proximity between susceptible people and infectors, an expressed limitation of the original model.

## INTRODUCTION

In today's society with increased urbanization, intense traveling and trading across the world, pathogens spread rapidly both within and between countries. Further, in times of COVID-19 and other pandemic or epidemic outbreaks as well as the rise of antibiotic resistance across the globe, humanity is in need of novel ways to understand pathogens. It is more important now than ever to understand how diseases are spread in different environments and learn how to prevent the transmission of infectious diseases.

The transmission of diseases in confined spaces is an especially interesting subject, since infected persons in need of treatment are concentrated in areas where a large number of people congregate such as in hospitals and their waiting areas. Furthermore, people with underlying conditions are common in these areas and they are particularly vulnerable to new infections that could become fatal. Infected individuals pose a risk for these vulnerable individuals. Higher crowd density has been consistently shown to increase infection risk.

Crowd simulation is an important tool in analyzing situations involving a large number of people or animals, where experimenting with actual subjects would be undesirable or impossible. There are several approaches, where agent- based models (ABM) are computational models designed to examine systems through the actions of autonomous agents. Recent research has shown that models currently in use can have their accuracy considerably improved. Here, accuracy refers to the model's output in comparison to real observations. A high accuracy would imply that the model is correct in its relationships be- tween variables in a dataset.

Tangentially, epidemiological models have been evaluated for the spread of tuberculosis in confined spaces and provide possible ways of comparing an agent-based implementation to earlier models and actual data. The GN (Gammaitoni and Nucci) model was found to be most accurate, but it is expressly limited by the fact that it does not account for the proximity of susceptible individuals to infectors. This could potentially be improved upon by an ABM. Tuberculosis is spread in a similar fashion as measles, which has a higher potential for airborne transmission in a space such as the waiting area of health care facilities.

Unlike some other epidemiological models, the GN equation includes the change of quanta level in a room space with time. Quanta is defined as the number of infectious droplet nuclei required to infect approximately 63,2% of susceptible persons when each susceptible person inhales one quantum of contagion . That definition becomes very useful in the comparison of different diseases as it creates an abstraction level where the precise infection mechanism is hidden and does not need to be accounted for.

## PROBLEM STATEMENT

When new viruses are emerging at the same time as there is an increase in urbanization, travelling and trading, viruses are easily spread both within and between countries. This makes it important to understand pathogens and how they are spread in different environments, particularly in environments where a large number of people are congregated and where people with underlying conditions are exposed. It is important in order to prevent transmission of infectious diseases that could turn fatal and at the same time facilitate and improve medical practices.

This paper aims to evaluate the feasibility of simulating an airborne transmission of infection in a confined space using an agent-based model. The epidemic model that will be used in the ABM (Agent-Based Model) is the Gammaitoni and Nucci model and the ABM will be implemented using the java-based MASON library core. The study will focus on transmissions of an airborne disease, specifically measles, in confined spaces. The viability of such an implementation will be evaluated by answering the research questions in section below.

This study could contribute towards the goal of properly understanding the spread of disease in confined spaces through simulation. The intrinsic value of such a simulation is to be able to further examine and manage infection risk without real-world subjects.

## RESEARCH QUESTIONS

**General Question**:
- Can the parameters of the GN (Gammaitoni- Nucci) model be implemented using an agent- based model to simulate the potential for airborne transmission of infection in a confined space?

By answering the following more specific questions regarding the agent-based model, we expect to be able to answer the general question with adequate de- tail. The answers can be found in section 5.

1. Is there a difference in infection rate when the volume of the confined space is altered, whilst ventilation rate and other parameters remain the same?
2. Is there a linear correlation between the density of the crowd in the con- fined space and the rate of infection?
3. Does the proximity of susceptible individuals to infectors change the average amount of quanta inhaled?

## APPROACH

Trials will be conducted using an ABM constructed by the authors with an existing java library called MASON. The model will consist of autonomous agents, of which some will be embodied and displayed as people. These virtual people will exist in a confined space and some will spread infectious droplets around them. The actual release of droplets will attempt to correspond to real data from previous research. The airborne disease will be absorbed by the embodied agents in the room, first by the ones who are closest to the person coughing and then by the rest of them as the droplets flow to the entirety of the room.

To answer the research questions, some parameters of the model will be altered in some crucial ways, such as the volume of the room and the number of agents. The weight of the agents as well as the placement of the agents will be taken into account when answering the questions. Primarily, it is desirable to mimic a real-life scenario to prove whether or not an ABM can be used to simulate the potential for airborne transmission of disease in a similar fashion to the GN model. The procedure here is to first analyze previous work and recreate it virtually. Then, it is possible to answer specific questions regarding crowd density and proximity to infectors through statistically comparing the number of infected individuals after a certain amount of time in different scenarios. After conducting these trials, there should be sufficient data to analyze whether or not key features of airborne infection risk are accounted for, following the line of questions posed in research questions.

# METHODS

## APPROACH

The construction of an ABM is a task which requires the designers to be un- biased as they are creating a reality for their agents to interact in. That reality must not be skewed in any direction as there are many complexities in analyzing the model's output [19] which could conceal the already unapparent bias. The approach therefore has its footing in simplicity and transparency, only implementing the features that are necessary and can be clearly evaluated according to the scientific method. Those features are chosen by looking at the parameters for the GN model and will be explained below.

## IMPLEMENTATION

### Time

The agents experience time in steps, where each step is configured to correspond to one second. Every new calculation is done when the simulation is "stepped". As all agents are autonomous we are faced with the question: in which order should they be stepped? If a person always breathes before the person next to them, they might absorb more particles from the air than their neighbor. Therefore, all people in the simulation are stepped in a random sequence, which means that their respective impact on the world is normally distributed. Furthermore, the ventilation of a room cannot be presumed to be constant in as small of a time fragment as a second, so it is also randomly inserted to be stepped with the people and normally distributed.

### Randomness

The ABM is built using the java language which comes with its own function to generate random numbers. This function is a linear congruential generator and suffers defects such as a short period and uneven distribution. In this simulation environment where randomness is key, it is insufficient. That is why all calls to generate random numbers go to an implementation of Mersenne Twister which returns uniformly distributed pseudorandom numbers with a period of $2^{19937} - 1$.

### The Environment

The embodied agents occupy space in a simulated hospital waiting room which is seen in figure 3.1. The environment has the following default parameters:
- There is exactly one infectious person present at all times
- The room is $132m^3$, length 8m, width 6m and height 2.75m
- The ventilation rate is 4 air changes per hour
- The number of susceptible people is 19
- The time spent in the room is 30 minutes

This is based upon a stochastic analysis done on the GN model for a waiting area of healthcare premises . The persons in the room are randomly seated on one of 35 seats.

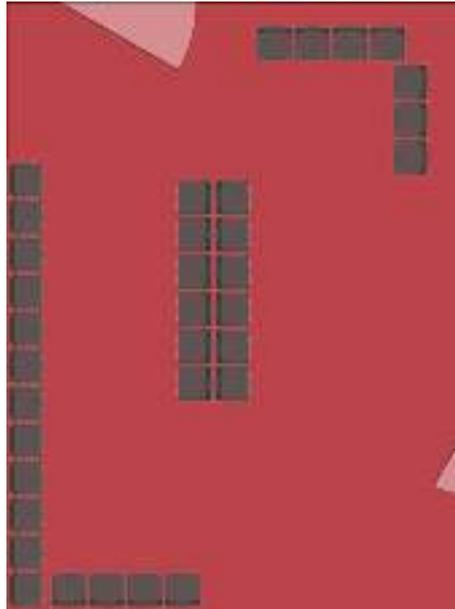
*A waiting room: the environment for the model*

## Space and Movement

The embodied agents are able to move about the room, and are not restricted to any type of grid. Instead they exist in a continuous two-dimensional space. This means that any and all distances are calculated with coordinates just as they would be in the real world. Collision detection was implemented for realism, however it is irrelevant to the purpose of the study and therefore agents by default pass through each other should they ever collide. Experiments were conducted with two, randomly chosen, persons moving and the rest sitting still in their seats. The two persons moving are connected with a band which can be seen in figure below.

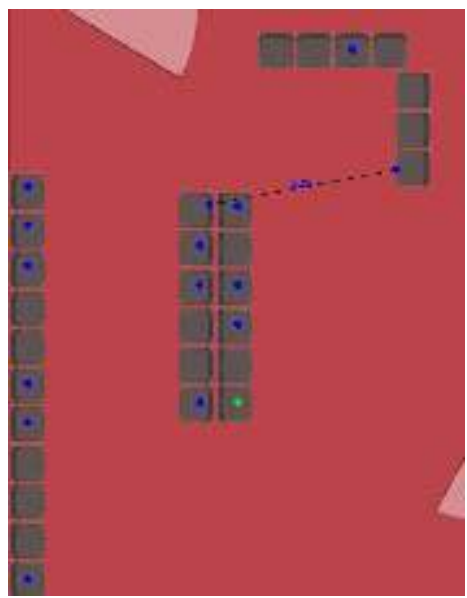
*Agents placed on seats in the waiting room. The blue dots represent susceptible agents, and the green dot represents the infector.*

The waiting room is divided into two volumes:
1. A cylinder with a base radius of 1.3m and height from the floor to the ceiling which is by default 2.75m. Will be referred to as the "cough area", which can be seen in figure 4.1.
2. The rest of the room.

This was done to be able to investigate the effect of proximity to an infectious person. The infector adds infectious droplets to the air in the room through coughing. Those droplets are first mixed with the air in the cylinder, which is within two seats from the infector, and subsequently flow to the rest of the room in which the air is well-mixed.

### Coughing

The airborne transmission of measles is the most likely to occur and that in turn happens through the infector coughing or sneezing. As it stands in the GN model, the method through which the virus becomes aerosolized is irrelevant. Instead, what is of importance is the quanta production rate. The measles virus has a mean quanta production rate of $570.0 q/h$ with a standard deviation of $143.0 q/h$. This production is simulated through the infector coughing every ten seconds and with $570.0 q/h = 570'000.0 mq/h$ and $143.0 q/h = 143'000.0 mq/h$ the quanta release per cough becomes $1583,33 \pm 397,2 mq$ and is randomly distributed.

In this research, there will be no distinction made between the droplet sizes and their respective modes of transmission.

### Breathing and Mass

People with a higher mass have a higher pulmonary ventilation. In other words, larger people take larger breaths. As they breathe more air pollutants, it is imperative to the purpose of this study to introduce people with different weights. The embodied agents are randomly assigned a weight at start and breathe according to their size. The largest assigned breath volume is 30% larger than the smallest.

When a person breathes, they absorb a volume of air that contains the aerosolized virus. That amount of virus is then removed from the mix of air in the space that they are in. These people will not produce any quanta of their own, as the simulation is not meant to run for long enough that the subjects get sick and start to produce more infectious droplets.

### Ventilation

The room is ventilated at a default rate of 4 air changes per hour. The air flow per second is calculated as

$$Airpersec = ACperhour * RoomVolume/(60*60).$$

That volume of air has a direct relation to the volume of the room, and that relation equals the percentage of the total quanta in the air-mix that is removed. This is because the air is well-mixed.

## EXPERIMENTS

The experiments were conducted by running the constructed ABM for 1800 steps which represents 30 minutes in time and by changing two parameters of the model; the volume of the room and the number of agents in the room. When changing the volume of the room, the height of the room was altered but the ventilation air flow remained static. Four different volumes of the room were tested, and the different amounts of agents were tested and compared. For each different scenario 10 or 20 simulations were conducted, partly due to time constraints but also due to the fact that a clear pattern quickly emerged.

The Figure below shows the control console of the ABM and what is shown when clicking on an agent in the waiting room. Under the tab "Inspectors" all properties of the agent are shown, such as the mass of the agent, total breath volume and amount of quanta inhaled which is the "infection" property. The mass property as well as the amount of quanta inhaled were the two properties that was examined in the experiments.

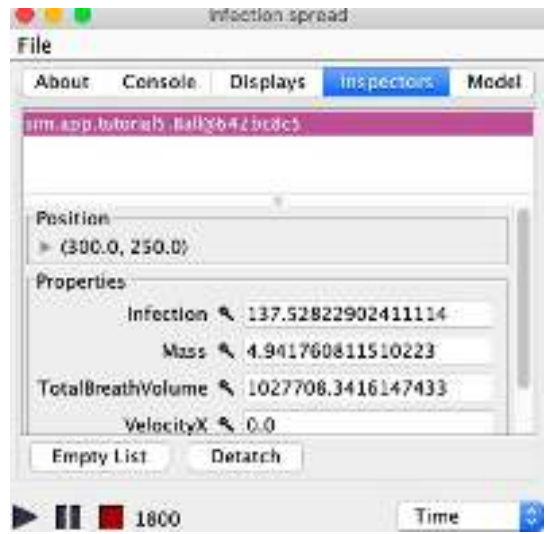

*The control console of the ABM.*

### Weight and Placement

To evaluate if the proximity of susceptible individuals to the infected person change the average amount of quanta inhaled we compared the infection property of agents sitting close to the infected agent with agents that sat further away. This was done by running simulations and retrieving the value of the infection property of each agent. The placement in relation to the infected agent as well as the weight of the agent (weights were divided into four different weight classes, further explained in the result section) were noted in a spreadsheet. For this scenario the default values mentioned above were used. In order to reach a fair amount of agents in each weight class and area, simulations were done until there were twenty agents per weight class and placement. This was because the probability of an agent being placed close to the infected agent was low, and much higher to be placed anywhere else in the room, away from the infector. Thereafter a mean value was calculated.

### Agent Density

In order to evaluate if there was a linear correlation between the density of the crowd in the confined space and the rate of infection, the number of agents in the room was altered. For these scenarios the default values of the environment were used for all other parameters. Three different amounts of agents were tested and ten simulations were run for each scenario. A mean value of the amount of quanta inhaled of each agent was calculated, but for each weight class, as described above. This was done for each simulation and the placement of the agents was not taken into account in these scenarios. The mean value for each weight class of ten simulations was calculated for all three scenarios.

### Volume of the Room

To evaluate if there is a difference in infection rate when the volume of the room is altered whilst ventilation rate and other parameters remain the same, experiments were conducted with the height of the room altered. The ventilation rate was a fixed value of air flow which was always equal to the default value of the room, $132m^3$ changed four times per hour. That equals $528m^3$ per hour, $8.8m^3$ per minute or $0.14667m^3$ per second. Four different heights were considered which implies four different volumes of the room. Furthermore, the number of agents was altered three times to see if it had any impact on the result. The simulation was run twenty times for each volume of the room and each amount of agents. The total amount of quanta in the room from each simulation was retrieved and a mean value was calculated from the data that was collected.

# RESULTS

The results of the simulation are found below and were produced in accordance with section 3.1. There will be no stochastic analysis of how many people fell ill in each scenario, instead focusing on the clearest information producible: the amount of quanta absorbed by each embodied agent. That amount will henceforth be referred to as the absorption amount and will be addressed further in the discussion section.

All values presented are after 1800 steps in the ABM, representing 30 minutes in the waiting room. The default parameters of the environment, found in section above on enviroment, applies to every simulation except when otherwise is stated. The weights that the agents could have were divided into four different weight classes with equal distributions. These are the classes that are referred to throughout this chapter:

- Weight class 1: Mass in 0-25% of distribution (Normal weight)
- Weight class 2: Mass in 25-50% of distribution (Overweight)
- Weight class 3: Mass in 50-75% of distribution (Obese class 1-2)
- Weight class 4: Mass in 75-100% of distribution (Obese class 2-3)

## The Impact of Placement

Depending on where the agents sat in the room in relation to the infected agent, which is the agent that was coughing, the absorption amount of each agent varied. A susceptible agent that sat within the "cough area" had a higher absorption amount than an agent that sat outside the cough area. Further, the absorption amount of each agent depended on the agents weight, in other words the mass property of the agent. During these simulations the default parameters of the environment were used meaning twenty agents were present.

Table 4.1: *Average inhaled infection (milliquanta for each weight class de- pending on the agents placement in relation to the infected agent.*

| *Weight class:* | *1* | *2* | *3* | *4* |
|---|---|---|---|---|
| *Outside cough* | 114,30 | 123,95 | 132,57 | 141,65 |
| *Inside cough* | 125,89 | 134,07 | 142,05 | 151,19 |

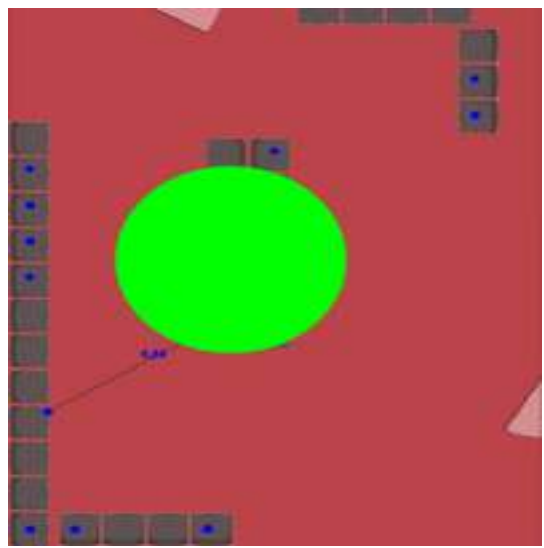

Figure 4.1: *Agent coughing in a simulation, the infected agent sits in the middle of the circle. Green circle: "Cough area".*

The table above presents the average inhaled infectious droplet amount in milliquanta for each weight class. Each average value is calculated from twenty agents per weight class from different simulations. "Inside cough area" is defined as being inside a cylinder centered around the infector, roughly within two chairs distance from the infected agents position, in other words the distance that the visualized cough (see figure 4.1) reaches in the ABM, shown above. "Outside cough area" is the remaining area in the waiting room. All values are presented in milliquanta. The values from table 4.1 are also presented in a chart (figure 4.2) below.

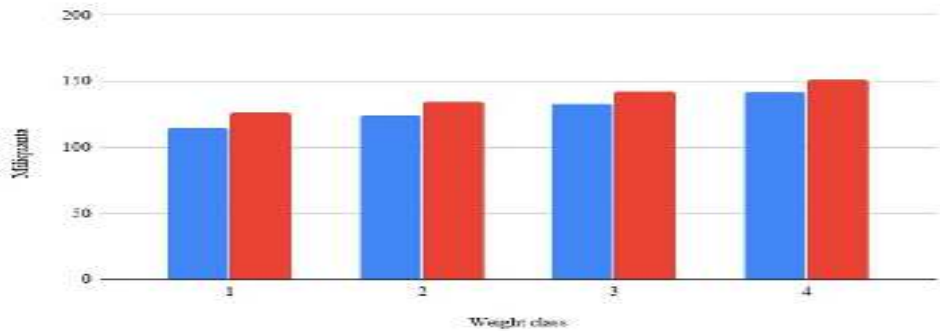

**Figure 4.2:** *Chart: Average absorption amount (milliquanta) for each weight class depending on the agents placement in the room in relation to the infected agent.*

The blue bars represent the average absorption amount per weight class when the agents sit far away from the infected agent, that is, outside the cough area. The red bars represents the average absorption amount per weight class for agents sitting close to the infected agent, that is, within two chairs away from the infected agents, in the cough area. All weight classes had a higher absorption amount when seated close to the infector.

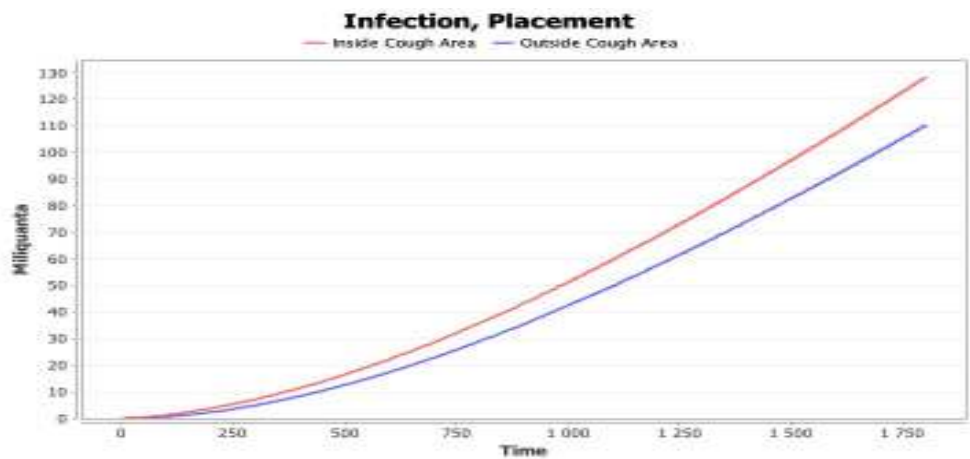

**Figure 4.3:** *Chart: Absorption amount for two different susceptible agents in the same weight class during 30 minutes depending on the agents placement in the waiting room.*

The absorption amount varies depending on the distance between the susceptible agent and the infected agent. Figure 4.3 represents two susceptible agents in the same weight class (weight class 1) in one simulation, and their inhalation of infectious droplets during 30 minutes in the waiting room. The red graph represents the agent that is sitting close to the infected agent, within the cough area, and the blue graph represents an agent that sits far away from the infected agent, outside the cough area.

## The Impact of Weight and Agent Density

As mentioned, people with a higher mass have a higher pulmonary ventilation. Therefore it was decided to investigate how much the ab- sorption amount varied in the different weight classes. The data presented in each table is the average absorption amount per weight class from ten different simulations. In addition to looking at the impact of weight, the experiments were repeated to examine the effect of crowd density. The number of agents present in the waiting area was altered three times throughout these simulations and were 10, 20 and 35 (including the infected agent). The amount of infectors was never changed, it was always one.

Table 4.2: *Average absorption amount per weight class. Agents: 10.*

| Weight Class | Milliquanta |
|---|---|
| 1 | 114,28 |
| 2 | 123,19 |
| 3 | 132,52 |
| 4 | 142,09 |

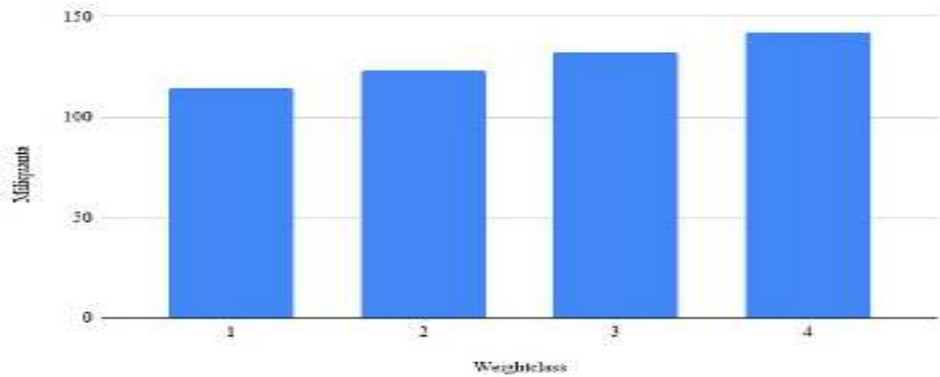

**Figure 4.4:** *Chart: Average absorption amount per weight class. Agents: 10.*

Results from when ten agents were present in the room are presented in table 4.2 and as a chart in figure 4.4. As stated earlier, the mean value is calculated from ten different simulations and the default values of the parameters of the room were used.

Table 4.3: *Average infection inhaled per weight class. Agents: 20.*

| Weight class | Milliquanta |
|---|---|
| 1 | 116,61 |
| 2 | 124,56 |
| 3 | 133,11 |
| 4 | 140,44 |

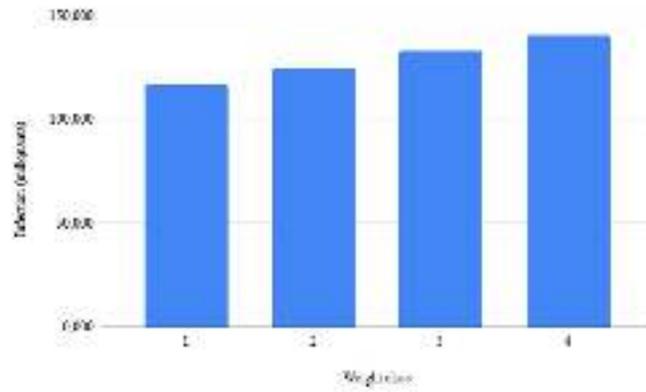

**Figure 4.5:** *Chart: Average absorption amount per weight class. Agents: 20.*

The results from when twenty agents were present in the room are presented in table 4.3 and as a chart in figure 4.5. The results from when 35 agents were present in the room are presented in table 4.4 and as a chart in figure 4.6.

Table 4.4: *Average absorption amount per weight class. Agents: 35.*

| Weight Class | Miliquanta |
|---|---|
| 1 | 116.28 |
| 2 | 121.32 |
| 3 | 132.51 |
| 4 | 139.56 |

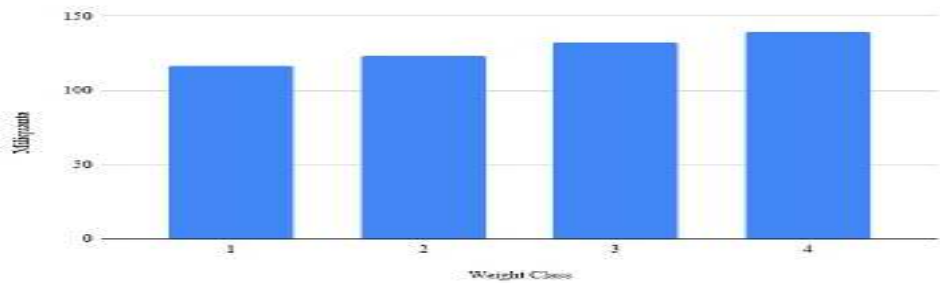

**Figure 4.6:** *Average absorption amount per weight class. Agents: 35.*

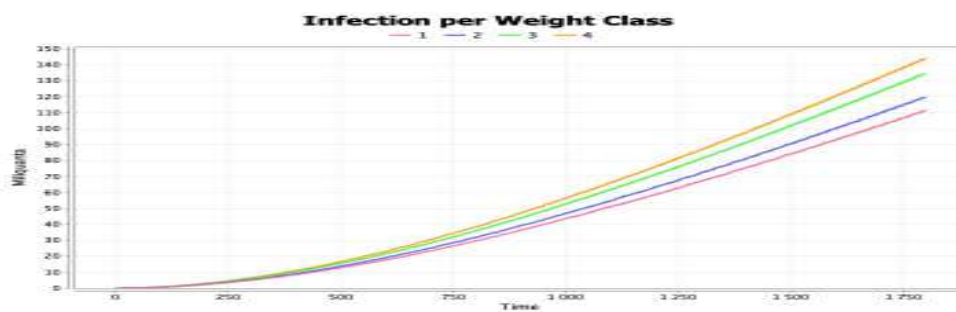

**Figure 4.7:** *Graph: Absorption amount during 30 minutes of four random agents in different weight classes.*

Figure 4.7 presents four random agents and their inhaled infection during 30 minutes in the waiting room. All agents belong to different weight classes. The red graph represents an agent in weight class 1, the blue graph represents an agent in weight class 2, the green graph represents an agent in weight class 3 and the orange graph represents an agent in weight class 4. This graph does not represent the results from all density tests, it is meant to visualize the difference in absorption amount between agents in different weight classes.

### 4.1.3 The Impact of the Room Volume and Agent Density

In these simulations the height of the room was altered, as well as the number of agents present in the room. One parameter was unaltered, the ventilation rate, which remained at the same air flow as in the default scenario. Four dif- ferent heights were tested; 2,40m, 2,75m, 3,5m and 5m. The different room volumes were therefore: $115,2m^3$, $132m^3$, $168m^3$ and $240m^3$. Twenty simulations were done for each volume and for each time the number of agents was altered, respectively. Simulations were done with 10, 20 and 35 agents present in the waiting room. The total amount of quanta in the room varied with the size of the room but not with the number of agents in the room which can be seen in figure 4.8 and table 4.5.

Table 4.5: *Average amount of quanta in the room (milliquanta), various volumes of the room and number of agents.*

| *Volume\Number of* | 10 | 20 | 35 |
|---|---|---|---|
| *115,2*m³ | 110949,02 | 111591,09 | 110302,44 |
| *132*m³ | 122433,70 | 122223,58 | 121923,36 |
| *168*m³ | 143748,64 | 142931,18 | 142352,72 |
| *240*m³ | 171056,40 | 170773,36 | 171302,66 |

**As seen in table 4.6 the concentration of quanta in the room decreased when the volume of the room increased.**

Table 4.6: *Average concentration of quanta in the room (milliquanta / $m^3$), various number of agents and volumes of the room.*

| Volume/Number of Agents | 10 | 20 | 35 |
|---|---|---|---|
| 115,2m³ | 963,10 | 968,67 | 957,49 |
| 132m³ | 927,53 | 925,94 | 923,66 |
| 168m³ | 855,65 | 850,78 | 847,34 |
| 240m³ | 712,74 | 711,56 | 713,76 |

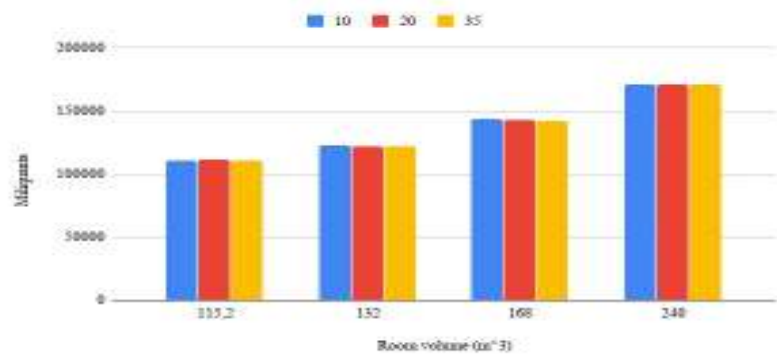

Figure 4.8: *Chart: Amount of quanta in the room with various heights of the room, and various number of agents.*

Figure 4.8 is the values from table 4.5 presented as a chart. The blue bars in the figure shows the amount of quanta in the room for each volume with 10 agents in the room. The red bar is the amount of quanta for each volume of the room with 20 agents and the yellow bar shows the amount of quanta in the room with 35 agents.

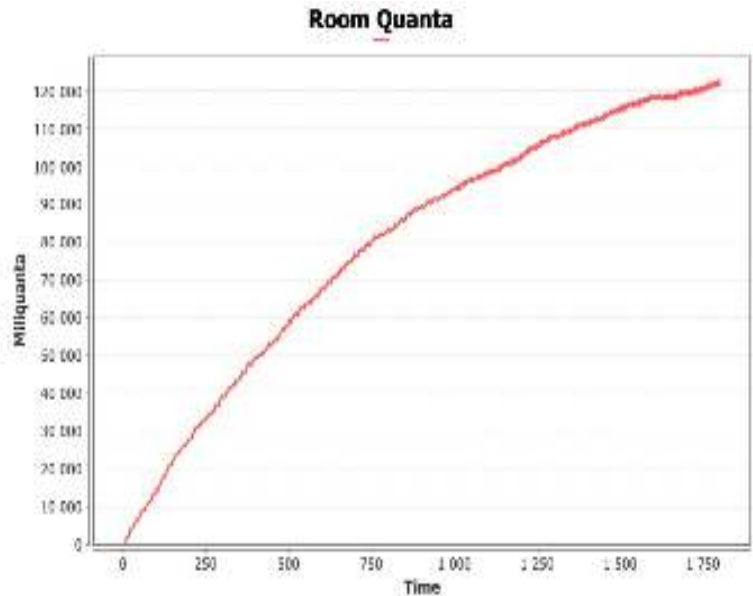

**Figure 4.9: Graph: Quanta in the room over time. Agents: 20. Height of the room: 2,75m.**

In figure 4.9 the total amount of quanta in the room is graphed during 30 min- utes. This graph is from one simulation with twenty agents present and with the default parameters of the environment. It is an example to showcase the overall result of total amount of quanta in the room.

# DISCUSSION
## Method

It was realized early on that building the model was going to take a significant amount of time and we therefore had to make the readings of its results as unambiguous as possible. Researching the amount of infectious droplets inhaled in units of quanta became a realistic goal as it did not require an extra level of medical or stochastic analysis, the raw data instead displaying a rather clear picture. The disadvantage of this method is that we cannot immediately draw conclusions regarding how many of the susceptible persons that would fall sick in a scenario like the one simulated. On the other hand, the comparisons between absorption amounts elucidate the impact of factors like proximity, mass and density. The overall lessened complexity proved it- self beneficial also while deciding the amount of simulations to run for each experiment as patterns rapidly emerged in the output. This allowed us to stay within the time frame while simultaneously remaining confident in the result.

Undoubtedly, the quality of the ABM implementation will be reflected in its output. The top priority when constructing it was to only insert simple functionality that in a system would simulate complex behavior. This is accounted for in section 3 but the inherent bias should be further discussed, as a computational model like this one is based on a set of rules chosen by the designer. In this case, those rules find their origin in previous peer-reviewed research but must be adapted creatively to answer the research question. As we are aware of the result of the previous research, our understanding of the topic and the accompanying hypotheses influence all creative decisions made. That is why effort was made to keep the amount of those decisions to an absolute mini- mum. It is also why it might be reassuring to implementers of ABMs to find results that contradict their hypotheses.

One specific design decision had to be made regarding the layout of the waiting room, the environment for the model. As reported in 3.2.3, parameters were taken from an earlier piece of research for ease of comparison, such as room length, width and height. However, to the model used in that research proximity did not matter and as such, the designs did not specify where furniture was placed or how people were expected to move. In coherence with the method approach, we wanted that decision to have as small of an impact as possible and chose an uncomplicated design with minimal movement. Plausibly, a model that added true to life movement would also have to investigate its effect on air mixing and the isolation of the factors we were interested in would have another dimension of complexity.

In conclusion, the scope of the paper meant limiting the features of the ABM. This was by all means necessary, and the scope itself was expected to become more narrow. That was the case as we eliminated any stochastic analysis from the project and focused instead on the three specific research questions. From the perspective of an ordinary citizen, the results would be more interesting if that analysis had been performed. Nevertheless, the results prove an important hypothesis in showing the viability of ABMs as alternatives to traditional epidemiological models. We believe that justifies the limitations, which in their own right made the research feasible.

## Result

As seen, the proximity to an infected person has an impact on how much quanta a susceptible person inhales according to the constructed ABM. It differs approximately 10 milliquanta between an agent that sits close to the infected agent and an agent that sits far away from the infected agent. That corresponds to 7-10 percentages depending on the weight class, where approximately 10 percentages is for weight class one and approximately 7 per- centages is for weight class 4. From these calculations it appears that the distance is of greater importance for someone with lower mass according to the ABM. However, we make no such assertion as the difference in percentages may not be statistically significant and there is no background in this paper to support such a statement.

Due to the shape of the graphs in figure 4.3, one can interpret that the effect of placement the first five minutes in the waiting room is negligible. The absorption amount remains under 10 milliquanta for both the agent within and the agent outside the cough area. The agent within the cough area will eventually consume a higher dose, but that pattern does not emerge clearly until around time step 300.

The construction of the room resulted in it being divided into two volumes; the "cough area" and the rest of the room. Since the infectious droplets the infector spread in the room were first mixed with the air in the cough area and subsequently flowed to the rest of the room, the concentration of quanta in the cough area was higher. This does not immediately replicate how it works in real life, but it was an approach to easily simulate that the concentration varies depending on the distance between an infected and a susceptible per- son. A more realistic simulation would have several more volumes with flow that eventually reaches the whole room, and a perfect one would instead simulate the flow of particles through the air. We are satisfied with the simple approach as it is adequate to investigate the proximity factor.

Another parameter that was crucial for the absorption amount is the weight of the agent, which is clearly seen in the data presented in table 4.1 and visualized in figure 4.2. This is due to the fact, as aforementioned, that larger people take larger breaths. That is also the way that agents in the ABM were constructed and, further, resulted in the output we expected. The results further strengthens the relationship between weight and absorption amount, where the data presented clearly shows that a susceptible with higher mass also has a higher absorption amount. The mean values for each weight class were calculated from all agents present in the simulation regardless of where they were sitting, since the mass property was the only parameter we focused on in these experiments. Although the position of the agents were not taken into consideration in the experiments conducted in this section, the average absorption amount per weight class presented in section 4.1.2 does not differ significantly from the data presented in the first row "outside cough area" in table 4.1. This was due to the fact that we did the simulations a sufficient amount of times to make the other parameters negligible when averaging the results.

What could also be seen, was that the density of the agents did not have any impact on the outcome of the results. This is possibly due to the fact that there was only one infector present in the room in all experiments that were conducted, which also means that the same amount of quanta was released in each simulation. This result was not expected since we thought there was a correlation between the density of the agents and the rate of infection. The only impact the density had was that more agents were now breathing in the infectious droplets. It had no apparent effect on the amount of quanta they absorbed, since the average absorption amount remained the same which could be seen in table 4.2-4.4 and figure 4.4-4.6. On the other hand, the results are reasonable since this was the way the ABM was constructed. All other parameters remained the same including the concentration of quanta in the room, which further implies that the amount of quanta inhaled per person remains largely the same and did not change as the number of agents increased. It should be noted that the absorption amount of one agent in comparison to the amount of quanta in the room is minuscule but existent and may have an effect if the number of agents is increased by a larger factor.

Furthermore, the volume of the confined space had an impact on the concentration of quanta in the room when the ventilation rate remained the same. The larger the room, the lower the concentration, which is clearly seen in table 4.5 and 4.6 as well as their associated chart in figure 4.8. It is most certainly an effect of that the same amount of quanta was released in a room with a different volume. What could be also seen is that the total amount of quanta in the room was lowered when the volume of the room was decreased, whilst the concentration was increasing at the same time. Additionally, this means that the individual absorption amount is lowered in higher volume rooms and in turn presumably also the infection rate.

It is clear from the data that the time aspect has an impact on the overall infection rate. If a susceptible person is present in a confined space with an infector of measles for a short time, the amount of quanta absorbed is negligible. This could be seen in figure 4.3, figure 4.7 and in figure 4.9, where it is clear that neither proximity nor weight has a significant impact on the absorption amount of each agent during the first six minutes spent in the room. This further shows the importance of short waiting times in waiting areas in order to reduce the risk of spreading diseases among people.

The division of weight classes could be done into even more classes which may have shown the impact of weight in a more distinguishable manner. Nonetheless, our division was reasonable for our time frame and resulted in a clear pattern, sufficiently clear for us to draw conclusions. Additionally, if there is a pattern for four different classes, that general pattern would exist for ten different classes as well, albeit with more local detail.

## FUTURE WORK

There are certain promising paths that lay ahead of this research and they will be presented as concisely as possible. Possible next steps to take in order to contribute to this area of research would be:

- To perform a stochastic analysis on the result of this thesis in order to further validate it and open up for a broader comparison to earlier work. Particularly interesting would be its relation to epidemiological models such as the SEIR- or Wells-Riley model.

- To add to the simulation different methods of transmission, such as person- to-person contact and contaminated objects. This would allow for a more complete transmission model and the possibility of investigating numerous diseases. In addition, a movement factor could be isolated and examined.

- To research the outcome if there were more than one infector in the room. This would support a more rigorous examination of crowd density as a factor in disease spread.

- To allow embodied agents to arrive and leave over a larger time span in order to demonstrate a more realistic hospital waiting area. This could provide more insights as to which factors, like ventilation or room vol- ume, have an impact on keeping the patients healthy.

# CONCLUSIONS

This study evaluated the feasibility of simulating an airborne transmission of infection in a confined space using an agent-based model and the parameters of the GN model. Further, the aim of the study was to find out if such an ABM could contribute to properly understand the spread of diseases in con- fined spaces without real-world subjects. The results of our research imply that the parameters of the GN model can be implemented using an agent-based model to simulate the potential for airborne transmission of infection in a con- fined space.

The data collected in the result section implies that the total amount of quanta in the room increases with the volume of the room, but the concentration of quanta in the room decreases, provided that the air flow to and from the room is unaltered. We have come to the conclusion that there is a difference in in- fection rate when the volume of the room is altered, whilst ventilation rate and other parameters remain the same, as the concentration is lowered and therefore also the individual absorption amount. Larger rooms lead to lower absorption amounts, answering research question 1. Further, with reference to research question 2, we do not draw any conclusion regarding a linear correlation between the density of the crowd in the confined space and the rate of infection when the number of infectors is static. As long as there only is one infector in the room it does not seem to matter how crowded the confined space is. No conclusions can be drawn from this research regarding a situation where the number of infectors change with the crowd density. Finally, concerning research question 3, the proximity of a susceptible individual to an infector has an impact in the average amount of quanta inhaled by the susceptible host. According to the constructed ABM, the amount of quanta inhaled is larger if a susceptible person is sitting next to the infector than if they are sitting farther away.